\documentclass[article,aps,nofootinbib,showpacs,amsfonts,epsf]{revtex4}

\input{epsf.tex}

\newcommand{\E}{{\cal{E}}}

\newcommand{\s}{\sigma}

\renewcommand{\d}{{\rm d}}
\renewcommand{\a}{\alpha}

\newcommand{\be}{\begin{equation}}
\newcommand{\ee}{\end{equation}}
\newcommand{\bea}{\begin{eqnarray}}
\newcommand{\eea}{\end{eqnarray}}
\newcommand{\ba}{\begin{array}}
\newcommand{\ea}{\end{array}}
\def\J#1#2#3#4{{#1} {\bf #2}, #3 (#4)}

\def\PR{Phys. Rev.}
\def\PRL{Phys. Rev. Lett.}
\def\PTP{Prog. Theor. Phys.}
\def\APNY{Ann. Phys. (NY)}

\def\JMP{J. Math. Phys.}

\def\CQG{Class. Quantum Grav.}

\def\JPCS{J. Phys.: Conf. Ser.}

\def\PLA{Phys. Lett. A}

\begin{document}
\draft

\title{Remarks on the mass--angular momentum relations\\ for two extreme Kerr sources in equilibrium}

\author{I.~Cabrera--Munguia,$^\dag$ V.~S.~Manko,$^\dag$ E.~Ruiz$\,^\ddag$}
\address{$^\dag$Departamento de F\'\i sica, Centro de Investigaci\'on y de
Estudios Avanzados del IPN, A.P. 14-740, 07000 M\'exico D.F.,
Mexico\\$^\ddag$Instituto Universitario de F\'{i}sica
Fundamental y Matem\'aticas, Universidad de Salamanca, 37008 Salamanca, Spain}

\begin{abstract}
The general analysis of the relations between masses and angular momenta in the configurations composed of two balancing extremal Kerr particles is made on the basis of two exact solutions arising as extreme limits of the well--known double--Kerr spacetime. We show that the inequality $M^2\ge|J|$ characteristic of an isolated Kerr black hole is verified by all the extremal components of the Tomimatsu and Dietz--Hoenselaers solutions. At the same time, the inequality can be violated by the total masses and total angular momenta of these binary systems, and we identify all the cases when such violation occurs. \end{abstract}


\maketitle


\section{Introduction}

In a recent paper \cite{DOr}, Dain and Ortiz have applied a numerical method to study the mass--angular momentum relations in a non--stationary axisymmetric system of two rotating black holes of the Kerr type \cite{Ker}. Using the notion of ``momentary stationary'' data, they have been able to reduce their initial problem to the stationary one for two black holes and a strut (conical singularity) in between. This allowed them, in particular, to interpret the presence of an unremovable strut in a system of two identical extreme black holes as an evidence of the non--existence of equilibrium configurations involving such black--hole constituents. The general physical message of the paper \cite{DOr} is transparent and somewhat expected: during the evolution of an axisymmetric system of two rotating black holes subjected to the gravitational and spin--spin interactions the inequality $M^2\ge|J|$ initially verified by the total mass and total angular momentum of the system is preserved. Here the positiveness of masses of both components is also assumed, which excludes the possibility of formation of binary equilibrium configurations by the non--extreme black holes \cite{Hoe,MRu}. Moreover, in the case of two extreme Kerr constituents with positive masses the absence of equilibrium configurations is anticipated as well, as an extension of analytical work on the non--extremal binary black--hole configurations; however, to the best of our knowledge, a general non--existence statement for the extreme case has never been articulated.

Since the axisymmetric model considered in \cite{DOr} does not permit the balance of its constituents, it looks desirable to supplement Dain and Ortiz's ``numerical evidences'' of the absence of regular equilibrium solutions for two extreme black holes obtained in the symmetric case, with the general analysis of the mass--angular momentum relations in the {\it exact} solutions for two extreme Kerr sources in equilibrium which arise within the framework of the well--known double--Kerr spacetime \cite{KNe}. By performing such an analysis, one would be able to find out, on the one hand, whether the aforementioned inequality characterizing any single subextreme or extreme Kerr source is also valid for two balancing extreme Kerr sources, and, on the other hand, whether there exist equilibrium configurations with positive Komar masses \cite{Kom} of both extremal constituents. The non--existence of the latter configurations might be regarded as an analytical proof of Dain and Ortiz's conjecture concerning the extreme two--body equilibrium solutions. It is worth reminding that the historically first equilibrium configuration describing two balancing extreme Kerr particles was discovered by Tomimatsu \cite{Tom} more than a quarter of a century ago as a result of his study of the balance conditions in the double--Kerr spacetime. Later on, the form of the Ernst potential \cite{Ern} for Tomimatsu's solution was found by Hoenselaers \cite{Hoe}, who in addition gave a rigorous proof that the solution had at least one naked singularity off the symmetry axis. The balance conditions of the double--Kerr solution were solved completely in the extreme case by Dietz and Hoenselaers \cite{DHo}, and their result can be used as basis for working out the corresponding balancing solutions in the explicit form.

The present paper pursues several objectives inspired by the work of Dain and Ortiz. First, we would like to give a full mathematical description of the Tomimatsu and Dietz--Hoenselaers spacetimes for two extreme Kerr constituents in equilibrium. Such a description is certainly necessary because for the Tomimatsu solution only the form of its Ernst potential is known, while the corresponding metric functions have never been calculated; in the case of the Dietz--Hoenselaers equilibrium configuration the situation is even worse -- neither its Ernst potential nor the expressions of metric functions were given in \cite{DHo}. To amend the state of things, a concise representation for each metric will be obtained  in our paper. Second, we will carry out a detailed analysis of the mass--angular momentum relations in the Tomimatsu and Dietz--Hoenselaers solutions with the aid of the analytical formulas for the individual and total physical quantities. A curious result which will follow from our study is the verification of the inequality $M^2\ge|J|$ by all the extreme constituents separately, and violation of this inequality by the total masses and angular momenta in some configurations which we shall specify fully. A demonstration that two extreme Kerr constituents with positive masses cannot form an equilibrium configuration could be considered as the third objective of our paper, and the non--existence of such configurations, as will be seen below, is a natural consequence of the implementation of the previous two objectives.

\section{The Tomimatsu solution}

The first exact solution for two extreme Kerr particles in equilibrium which we are going to consider is the Tomimatsu solution \cite{Tom} which arises, within the framework of the double--Kerr solution \cite{KNe}, as a special limiting case of the balance conditions. The form of the Ernst potential defining this spacetime was given by Hoenselaers \cite{Hoe}:
\bea
\E&=&(A-B)/(A+B), \nonumber\\
A&=&2(x^2-y^2)^2-l(x^4+y^4-2)+2i[(x+y)^2(1-xy)+l(x-y)(2xy^2-x+y)], \nonumber\\
B&=&2\{l[x(x^2-1)-y(1-y^2)]-(x+y)(x^2-y^2)+il(x-y)[1-y^2-y(x+y)]\}, \label{Tom_E}  \eea
where $x$ and $y$ are prolate spheroidal coordinates, and $l$ is an arbitrary real constant. The relation of $x$ and $y$ to the Weyl--Papapetrou cylindrical coordinates $(\rho,z)$ is given by the formulas
\be
x=\frac{1}{2\a}(r_++r_-), \quad y=\frac{1}{2\a}(r_+-r_-), \quad r_\pm=\sqrt{\rho^2+(z\pm\a)^2}, \label{xy_rz} \ee
the positive real parameter $\a$ representing a coordinate distance from the location of each extreme Kerr constituent (the points $\rho=0$, $z=\pm\a$) at the symmetry $z$--axis to the origin of coordinates (see Fig.~1).

In the paper \cite{Hoe} the potential (\ref{Tom_E}), which is a subcase of the Kinnersley--Chitre 5--parameter solution \cite{KCh}, was used for proving that the Tomimatsu spacetime has at least one naked singularity off the symmetry axis due to the presence of a negative mass. However, apart from the Ernst potential, it is also advantageous for various reasons to know the corresponding metric functions $f$, $\gamma$ and $\omega$ entering the stationary axisymmetric line element
\be
\d s^2=\a^2f^{-1}\left[e^{2\gamma}(x^2-y^2)\left(\frac{\d x^2}{x^2-1}+\frac{\d y^2}{1-y^2}\right)+(x^2-1)(1-y^2)\d\varphi^2\right]-f(\d
t-\omega\d\varphi)^2, \label{Papa} \ee
so below we shall give the whole metric defined by the potential (\ref{Tom_E}).

Recall that the metric function $f$ is determined as the real part of $\E$: $f={\rm Re}(\E)$; the metric coefficient $\omega$ is obtainable from $\E$ by solving the differential equations
\be
\omega_{,x}=\frac{4\a(y^2-1){\rm Im}(\E_{,y})}{(\E+\bar\E)^2}, \quad \omega_{,y}=\frac{4\a(x^2-1){\rm Im}(\E_{,x})}{(\E+\bar\E)^2} \label{w_eq} \ee
(a bar over a symbol means complex conjugation), while the equations for $\gamma$ have the form
\bea
\gamma_{,x}=\frac{1-y^2}{(x^2-y^2)(\E+\bar\E)^2}[x(x^2-1)\E_{,x}\bar\E_{,x} -x(1-y^2)\E_{,y}\bar\E_{,y}-y(x^2-1)(\E_{,x}\bar\E_{,y}+\bar\E_{,x}\E_{,y})], \nonumber\\ \gamma_{,y}=\frac{x^2-1}{(x^2-y^2)(\E+\bar\E)^2}[y(x^2-1)\E_{,x}\bar\E_{,x} -y(1-y^2)\E_{,y}\bar\E_{,y}+x(1-y^2)(\E_{,x}\bar\E_{,y}+\bar\E_{,x}\E_{,y})]. \label{g_eq} \eea

In the case of the Tomimatsu solution we have been able to write down the final expressions for $f$, $\gamma$ and $\omega$ in terms of four basic polynomials $\mu$, $\sigma$, $\pi$, $\tau$:
\bea f&=&\frac{N}{D}, \quad e^{2\gamma}=\frac{N}{(l-2)^2(x^2-y^2)^4}, \quad \omega=\frac{\a(1-y^2)F}{N}, \nonumber\\ N&=&\mu^2-(x^2-1)(1-y^2)\s^2, \nonumber\\ D&=&N+\mu\pi-(1-y^2)\s\tau, \nonumber\\ F&=&(x^2-1)\s\pi-\mu\tau, \nonumber\\ \mu&=&(2-l)(x^2-y^2)^2+2l(x^2-1)(1-y^2), \nonumber\\ \s&=&2[(x+y)^2-2ly(x-y)], \nonumber\\ \pi&=&4(l-2)^{-1}\{2(x+y)^2(x-y-1)-l(x+y)[(x-y)^2+2(x-1)^2] \nonumber\\ &+&l^2[x(x-1)^2+y(1-y)^2]\}, \nonumber\\ \tau&=&4l(l-2)^{-1}(x-1)[(l-2)(x-y)(x+2y-1)-2(x+1)], \label{Tm} \eea
thus arriving at a representation of the metric functions which is analogous to the one proposed by Perj\'es~\cite{Per} for the well--known Tomimatsu--Sato spacetimes~\cite{TSa,Yam}. Mention that for finding the form of the polynomials $\mu$ and $\sigma$ we made use of the results of the paper~\cite{Yam2} on the Kinnersley--Chitre metric, while the form of $\omega$ was found by a straightforward integration of equations (\ref{w_eq}).

It can be easily seen that the function $\gamma$ vanishes on the symmetry axis everywhere outside the sources, and the numerator of the function $\omega$ has the factors $(1-y^2)$ and $(x-1)$, which means that the symmetry axis is regular on the intervals $z>\a$, $z<-\a$ and $|z|<\a$, i.e., the extreme Kerr sources are in equilibrium.

The total mass and total angular momentum of the system can be found by analyzing the asymptotic behavior of the metric (\ref{Tm}), yielding
\be
M_{\rm T}=\frac{2\a(l-1)}{2-l}, \quad J_{\rm T}=\frac{2\a^2(l^2-2)}{(2-l)^2}, \label{MJT_tot} \ee
and, as was already pointed out in \cite{Tom}, the positive values of $M_{\rm T}$ correspond to $1<l<2$.

The expressions for individual masses of the constituents have the form \cite{Tom}
\be
M_1=-\frac{2\a}{2-l}, \quad M_2=\frac{2\a l}{2-l}, \label{MT_ind} \ee
the subindices 1 and 2 referring to the upper and lower constituents, respectively, while the corresponding expressions for the angular momenta of the constituents can be worked out with the aid of the general formulas (3.8) of \cite{Tom}, yielding
\be
J_1=-\frac{4\a^2}{(2-l)^2}, \quad J_2=\frac{2\a^2l^2}{(2-l)^2}. \label{JT_ind} \ee
It is worth noting that the aforementioned formulas (3.8) of \cite{Tom} give for $J_1$ a positively defined quantity $4\a^2/(2-l)^2$; however, our choice of sign in the formula (\ref{JT_ind}) for $J_1$ is justified not only by the expression (\ref{MJT_tot}) for the total angular momentum, but also by an independent numerical check of the Komar quantities that we made for the Tomimatsu solution using the metric (\ref{Tm}) and integral formulas (23) of the paper \cite{MRS}. It is easy to see that the above $M_i$ and $J_i$ verify the equilibrium formula
\be
J_1+J_2+2\a\left(\frac{J_1}{M_1}+\frac{J_2}{M_2}\right)\pm(M_1+M_2+2\a)^2=0 \label{MR_law} \ee
(with the low sign) obtained in the paper \cite{MRu2}.

We now turn to the study of the mass--angular momentum relations in the Tomimatsu solution, and we will start with the total quantities, first considering the positive values of $M_{\rm T}$, defined by $1<l<2$, in order to compare our results with the analysis of Dain and Ortiz. Then from (\ref{MJT_tot}) follows that the inequalities
\be
M_{\rm T}^2\ge|J_{\rm T}|, \quad M_{\rm T}>0 \label{MgJ} \ee
are satisfied only at the interval $(4/3)\le l<2$, whereas for $1<l<(4/3)$ the inequalities verified by $M_{\rm T}$ and $J_{\rm T}$ are
\be
M_{\rm T}^2<|J_{\rm T}|, \quad M_{\rm T}>0. \label{MlJ} \ee

Therefore, we have arrived at an interesting conclusion that {\it there exist equilibrium configurations of two extremal Kerr particles possessing positive total mass for which the inequality} $M\ge\sqrt{|J|}$, {\it considered by Dain and Ortiz in \cite{DOr}, is not satisfied}. Apparently, this does not invalidate the analysis of the paper~\cite{DOr} because one of the masses in Tomimatsu's solution is negative.

From now on we will not restrict the masses to only positive values and will be interested in the validity of the inequality
\be
M^2\ge|J|  \label{MJ_gen} \ee
both for the total and individual Komar quantities of the Tomimatsu solution.

We note that $M_{\rm T}$ and $J_{\rm T}$ defined by (\ref{MJT_tot}) verify in general the relation
\be
M_{\rm T}^2-J_{\rm T}=2\a^2, \label{MJT_gr} \ee
which means that the inequality (\ref{MJ_gen}), with $M=M_{\rm T}$, $J=J_{\rm T}$, can be violated in principle only by the negative values of $J_{\rm T}$ determined by a narrow interval $|l|<\sqrt{2}$. However, a simple check shows that the inequality $M_{\rm T}^2\ge|J_{\rm T}|$ is verified by some negative values of $l$ too, so in reality the interval at which the latter inequality is violated reduces to $0<l<(4/3)$. In Fig.~2 we have plotted, for $\a=1$, the quantity $\Delta_{\rm T}\equiv M_{\rm T}^2-|J_{\rm T}|$ as a function of $l$ to illustrate the relation between $M_{\rm T}$ and $J_{\rm T}$ in the Tomimatsu solution.

Let us now discuss the individual masses and angular momenta of the extreme constituents. We first find from (\ref{MT_ind}) and (\ref{JT_ind}) that $M_1$ and $J_1$ satisfy the equality $M_1^2=|J_1|$, which means that the first constituent is endowed with the same mass--angular momentum relation as a single extremal Kerr source, no matter whether $M_1$ is positive or negative. On the other hand, the equality $M_2^2=|J_2|$ is only possible for $l=0$, in which case the Tomimatsu solution reduces to the usual extreme Kerr spacetime. When $l\ne0$, the quantities $M_2$ and $J_2$ verify the inequality $M_2^2>|J_2|$ which is characteristic of an isolated subextreme Kerr black hole. In Fig.~3 the values of $\Delta_{2}\equiv M_{2}^2-|J_{2}|$ are plotted as functions of $l$ for $\a=1$.

Therefore, we can conclude that the Tomimatsu solution represents a remarkable equilibrium configuration of two extreme Kerr sources in which the individual masses and angular momenta of the constituents satisfy the inequality (\ref{MJ_gen}) in general, but the total quantities $M_{\rm T}$ and $J_{\rm T}$ fail to do so for some values of the parameter $l$!

It is worth noting that the extreme constituents described by the solution (\ref{Tom_E}) are asymmetric in the sense that they cannot exchange characteristics with their partner. As a consequence, for instance, in the configurations defined by a positive total mass, the mass $M_1$ of the upper constituent is always a negative quantity, whereas the mass $M_2$ of the lower constituent is always a positive quantity. The solution in which the constituents of Tomimatsu's configurations interchange their places was pointed out by Dietz and Hoenselaers (see formula (4.31c) of \cite{DHo}), and its Ernst potential is obtainable from the potential (\ref{Tom_E}) via the substitution $y\to-y$ and subsequent complex conjugation. However, such a modified Tomimatsu's solution obviously cannot give us any new information about the physical properties of balancing extreme Kerr sources.

\section{The Dietz--Hoenselaers solution}

The second exact spacetime describing two extreme Kerr constituents in equilibrium is defined by the solution of the balance equations reported by Dietz and Hoenselaers in the paper \cite{DHo} (see formulas (4.31a) of \cite{DHo}) and arising directly as extreme limit of the general equilibrium formulas found by Hoenselaers for the subextreme case \cite{Hoe}. To construct the Ernst potential and the whole metric for this configuration of extreme Kerr sources which differs from the Tomimatsu spacetime considered in the previous section, we found it convenient to employ the general formulas of the extended double--Kerr equilibrium problem \cite{MRu2} which are more familiar to us. The procedure for performing the extreme limit ($\a_2=\a_1$, $\a_4=\a_3$) is outlined in \cite{CMR} where some preliminary results have been also obtained.\footnote{There is a misprint in formulas (5) of \cite{CMR}: the second term in the expression for $\Lambda_2$ should not have the factor 2.} Below we give the form of all the functions defining the Dietz--Hoenselaers field in a representation similar to the one we have already used for the Tomimatsu metric:
\bea \E&=&\frac{A-B}{A+B}, \quad f=\frac{N}{D}, \quad e^{2\gamma}=\frac{N}{(a+3)^2(x^2-y^2)^4}, \quad \omega=\frac{4\epsilon\a(x-1)(1-y^2)F}{(a+3)N}, \nonumber\\
A&=&(a+3)(x^2-y^2)^2-2(a-1)(x^2y^2-1)+2i\epsilon\{xy[(a+1)(x^2-1) \nonumber\\ &-&(3-a)(1-y^2)]+(x^2-2x^2y^2+y^2)\}, \nonumber\\
B&=&2\{-x[2(x^2-1)-(a^2-a-2)(1-y^2)]+by[(a+1)(x^2-1)+2(1-y^2)] \nonumber\\ &-&i\epsilon(a-1)[(a+2)y(x^2-1)-(bx-2y)(1-y^2)]\}, \nonumber\\
N&=&\mu^2-(x^2-1)(1-y^2)\s^2, \nonumber\\ D&=&N+\mu\pi-(1-y^2)\s\tau, \nonumber\\ F&=&(x+1)\s\pi-\mu\tau, \nonumber\\ \mu&=&(a+3)(x^2-y^2)^2+2(a-1)(x^2-1)(1-y^2), \nonumber\\ \s&=&2[(a+1)(x^2-y^2)+4y(y-bx)], \nonumber\\ \pi&=&4[(ax-by)^2+(bx-y)^2]-(a+3)\{2(x-by)(x^2-y^2) \nonumber\\ &+&(a-1)[ax(y^2+1)-by(x^2+1)]\}, \nonumber\\ \tau&=&(a-1)\{-(a+3)[(a+2)x^2-2y^2+by(x+1)] \nonumber\\
&-&(a-1)(a+2)x+2(a+1)\}, \label{DHm} \eea
where $\a$ is the same distance parameter as in the previous solution, and the constants $\epsilon$, $a$ and $b$ are defined as follows
\be
\epsilon=\pm1, \quad a=p+q, \quad b=p-q, \quad p^2+q^2=1. \label{abpq} \ee

On the upper part of the symmetry axis ($y=1$) the Ernst potential of the solution (\ref{DHm}) takes the form
\bea \E(\rho=0,z)&=&E_+/E_-, \nonumber \\ E_\pm&=&(a+3)z^{2}\pm 2\alpha[2\pm \mathrm{i}\epsilon(a+1)]z \nonumber\\ &-&\alpha^{2}\{1\pm 2b+a(3\pm 2b)\pm 2\mathrm{i}\epsilon[(a+2)(1-a)\pm 2b]\}, \label{DN_axis} \eea
whence the total mass and total angular momentum of the system can be easily obtained by calculating the first two Geroch--Hansen multipole moments \cite{Ger,Han} with the aid of the procedure \cite{FHP}, yielding
\be
M_{\rm T}=-\frac{4\a}{3+a}, \quad J_{\rm T}=\frac{2\epsilon\a^2[(2-a)(3+a)^2-8]}{(3+a)^2}. \label{DH_tot} \ee

From (\ref{DH_tot}) follows that the total mass of the Dietz--Hoenselaers solution can take only negative values.

The expressions for individual masses and angular momenta of the extreme constituents in the equilibrium configurations described by the metric (\ref{DHm}) can be readily worked out from the respective formulas of the paper \cite{MRu} by taking in them the limit $\a_2=\a_1=\a$, $\a_4=\a_3=-\a$ and making the redifinitions $p\to\epsilon_1 p$, $q\to\epsilon_4 q$:
\bea
M_1&=&\frac{2\a p(p-1)}{p+q+pq-1}, \quad M_2=\frac{2\a q(q-1)}{p+q+pq-1}, \nonumber\\ J_1&=&\frac{2\epsilon\a^2p^2(p-1)(p-q+pq-1)}{(p+q+pq-1)^2}, \quad J_2=\frac{2\epsilon\a^2q^2(q-1)(q-p+pq-1)}{(p+q+pq-1)^2}, \label{MiJiDH} \eea
and one can see that the above expressions are symmetric with respect to the change $p\to q$, $q\to p$, i.e., a constituent in this kind of configurations may exchange the characteristics with its partner. Apparently, the above $M_i$ and $J_i$ fulfil the equilibrium formula (\ref{MR_law}).

It is not difficult to show that, like in the case of the Tomimatsu solution, the majority of the equilibrium configurations described by the Dietz--Hoenselaers solution verify the inequality $M_{\rm T}^2>|J_{\rm T}|$ for the total mass and total angular momentum. Indeed, the sign of the quantity $\Delta_{\rm T}=M_{\rm T}^2-|J_{\rm T}|$ coincides with the sign of the expression
\be
1-p^3-q^3+4pq, \ee
so that the values of $p$ and $q$ for which $\Delta_{\rm T}>0$ are the following:
\bea
-1<p<p_-, \quad &&q>0 \quad \mbox{or} \quad q<0; \label{dg01} \\
p_-<p<0, \quad &&q<0; \label{dg02} \\
0<p<p_+, \quad &&q>0 \quad \mbox{or} \quad q<0; \label{dg03} \\
p_+<p<1, \quad &&q>0; \label{dg04} \\
p_\pm=\frac{1}{4}\Bigl(\sqrt{17}-5&\pm&\sqrt{10\sqrt{17}-34}\Bigr) \nonumber \eea
(up to three decimal places, $p_+\simeq0.453$ and $p_-\simeq-0.891$). In particular, the case of two identical corotating extreme Kerr sources defined by $p=q=\sqrt{2}/2$ arises within the parameter range (\ref{dg04}). On the other hand, $\Delta_{\rm T}$ is a negative quantity for the following $p$ and $q$:
\bea
&&p_-<p<0, \quad q>0; \label{dl01} \\
&&p_+<p<1, \quad q<0, \label{dl02} \eea
and these values differ from (\ref{dg02}) and (\ref{dg04}) only in the sign of the parameter $q$. No analogs of the intervals (\ref{dg01}) and (\ref{dg03}) appear in the case $\Delta_{\rm T}<0$.

As far as the individual masses are concerned, it is important to note that none of the constituents violates the inequality $M_i^2\ge|J_i|$. The positive values of $M_1$ are covered by $p$ and $q$ verifying
\be
0<p<1, \quad q<0, \label{m1g0} \ee
while the negative values of $M_1$ correspond to
\bea
-1<p<0, \quad &&q>0 \quad \mbox{or} \quad q<0; \nonumber\\
0<p<1, \quad &&q>0, \label{m1l0} \eea
and one can see that there are more opportunities for $M_1$ to take negative values than the positive ones.

Similarly, $M_2$ is positive for
\be
-1<p<0, \quad q>0, \label{m2g0} \ee
and negative for
\bea
0<p<1, \quad &&q>0 \quad \mbox{or} \quad q<0; \nonumber\\
-1<p<0, \quad &&q<0. \label{m2l0} \eea

From (\ref{m1g0}) and (\ref{m2g0}) follows that $M_1$ and $M_2$ cannot simultaneously take positive values; however, the case $M_1<0$, $M_2<0$ is allowed, and it arises when $p$ and $q$ verify
\bea
-1<p<0, \quad &&q<0; \nonumber\\
0<p<1, \quad &&q>0. \label{m1m2l0} \eea

Because of the presence of a negative mass in the Dietz--Hoenselaers solution, the latter has at least one massless ring singularity off the symmetry axis. The proof of this fact is based on the analysis of zeros of the real and imaginary parts of the function $A+B$, and is fully analogous to the one given by Hoenselaers \cite{Hoe} in the case of Tomimatsu's solution. Therefore, we do not give it here; instead, for illustrating that the naked singularity is developed by the constituent with a negative Komar mass, we have plotted in Fig.~4 the stationary limit surface ($f=0$) of the Dietz--Hoenselaers solution for a particular choice of the parameters $\a=1$, $p=3/5$, $q=-4/5$ leading to $M_1=2/7$ and $M_2=-12/7$. The naked singularity occurs at $\rho\simeq 1.337$, $z\simeq-0.929$ and lies on the stationary limit surface of the lower constituent possessing negative mass.

\section{Conclusions}

In the present paper we have obtained a concise representation of all metrical functions for two analytic solutions describing equilibrium configurations of a pair of extreme Kerr sources, and analyzed the relations between the masses and angular momenta in such binary systems. We have shown that neither in the case of the Tomimatsu solution nor in that of Dietz and Hoenselaers the inequality $M^2\ge|J|$ is violated by the individual masses and angular momenta. At the same time, for both solutions there exist configurations in which this inequality is violated by the total Komar quantities, the latter satisfying already the relation $M_{\rm T}^2<|J_{\rm T}|$. The violation, however, takes place exclusively in the configurations involving one positive and one negative mass.

Since the Tomimatsu and Dietz--Hoenselaers solutions cover all possible values which the masses of two extreme Kerr sources can take in an equilibrium configuration, from our analysis follows straight away that {\it the balance of two extreme Kerr constituents with positive masses is impossible}. This conclusion probably tells us that more attention should be paid in the future to the systems comprised of three black holes as potentially more promising for searching the physically meaningful equilibrium configurations. Indeed, it was already shown within the framework of the triple--Kerr solution \cite{MRM} that there exist special equilibrium states of three non--extremal Kerr black holes possessing positive Komar masses; therefore, one might expect that a clever combination of analytical approach based on an appropriate three--body exact solution and numerical methods, such as the one developed by Dain and Ortiz \cite{DOr}, would be able to shed additional light on the balance of rotating extreme black holes.

\section*{Acknowledgments}

This work was supported by Project FIS2006-05319 from Ministerio de Ciencia y Tecnolog\'\i a, Spain, and by the Junta de Castilla y Le\'on under the ``Programa de Financiaci\'on de la Actividad Investigadora del Grupo de Excelencia GR-234'', Spain.

\newpage

\begin{figure}[htb]
\centerline{\epsfysize=90mm\epsffile{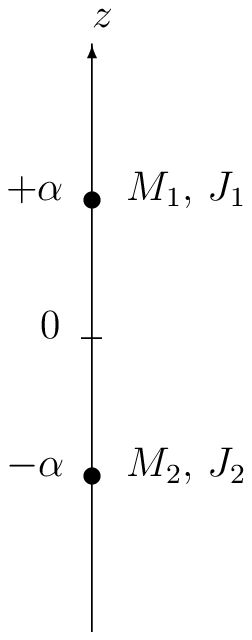}} \caption{Location of two extreme Kerr constituents on the symmetry axis.}
\end{figure}

\newpage

\begin{figure}[htb]
\centerline{\epsfysize=90mm\epsffile{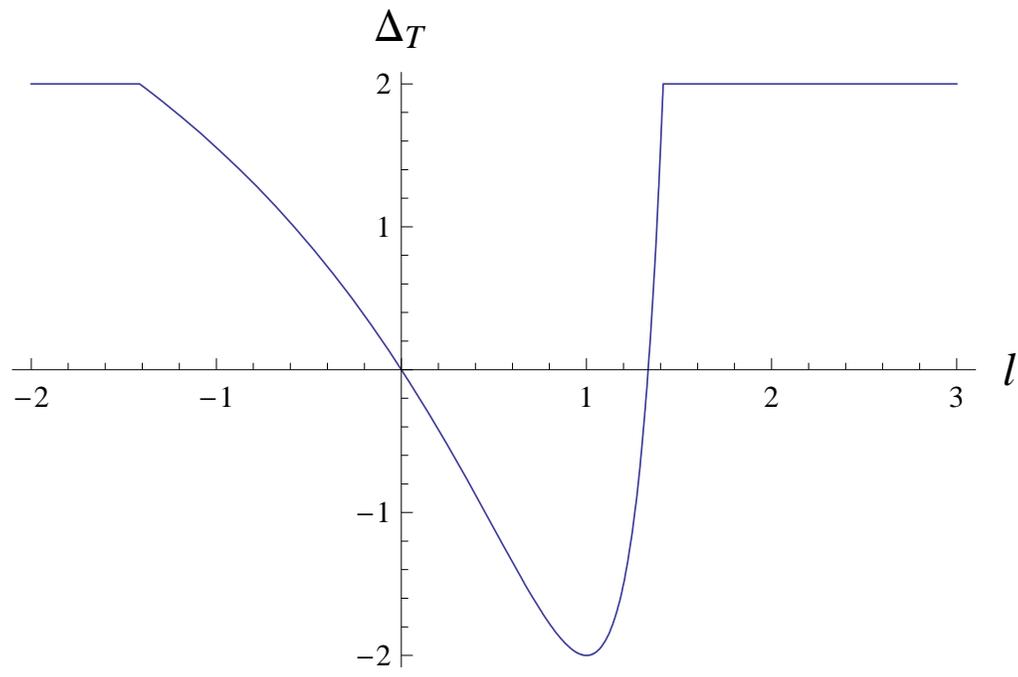}} \caption{The $\Delta_{\rm T}$ as a function of $l$ diagram for $\a=1$ in the case of the Tomimatsu solution. The curve intersects the $l$--axis at the points $l=0$ and $l=\textstyle\frac{4}{3}$.}
\end{figure}

\newpage

\begin{figure}[htb]
\centerline{\epsfysize=90mm\epsffile{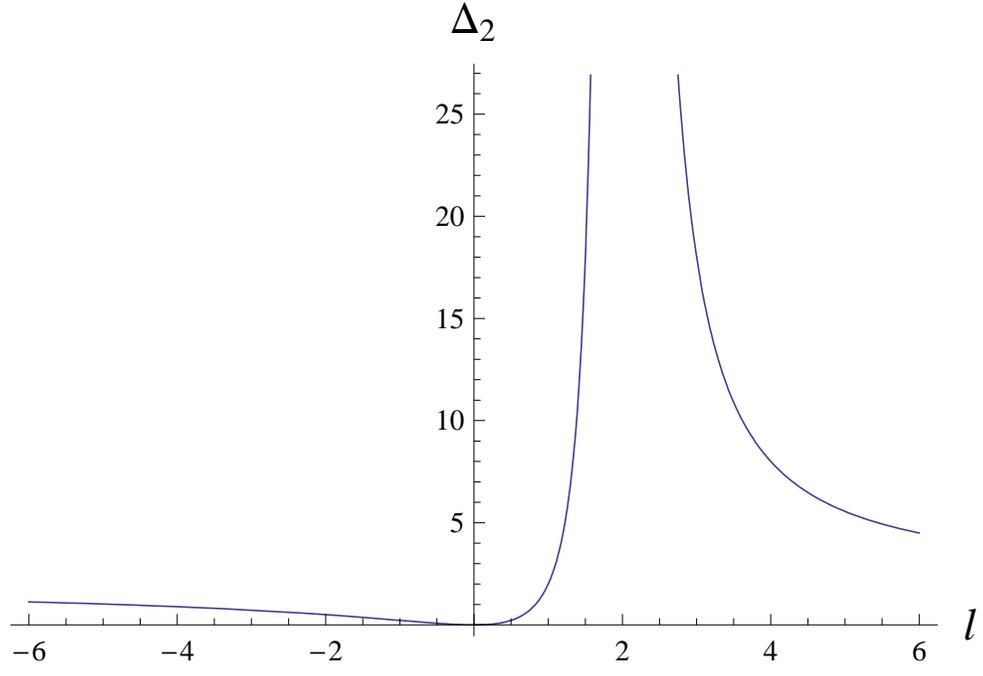}} \caption{The $\Delta_2$ as a function of $l$ diagram in the case of the Tomimatsu solution. The parameter $\a$ is set equal to unity.}
\end{figure}

\newpage

\begin{figure}[htb]
\centerline{\epsfysize=90mm\epsffile{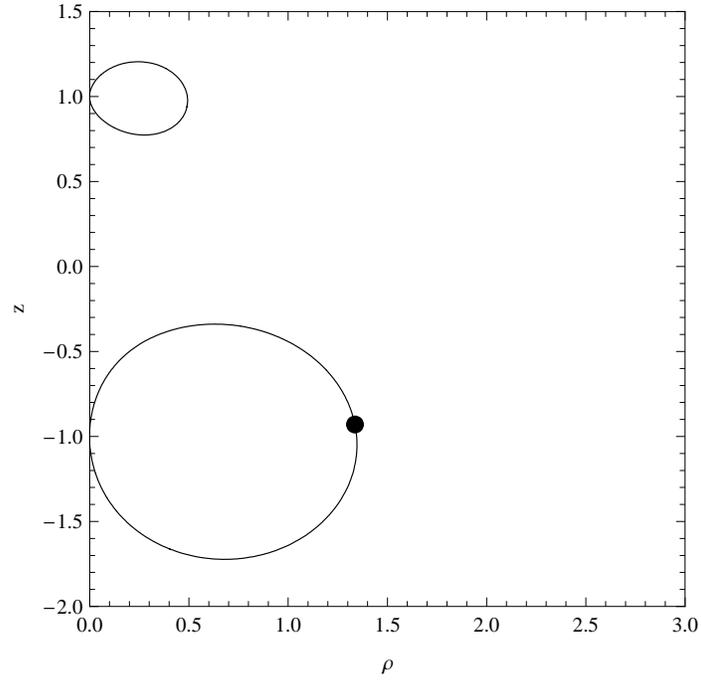}} \caption{Location of a naked singularity in the case of the Dietz--Hoenselaers solution. The parameters have been assigned the following values: $\a=1$, $p=3/5$, $q=-4/5$.}
\end{figure}

\end{document}